\documentclass[
    ,final            % use final for the camera ready runs
  ]
  {aipproc}

\layoutstyle{6x9}

\newcommand\aj{AJ}
\newcommand\apj{ApJ}
\newcommand\mnras{MNRAS}

\begin{document}

\title{Dark matter and the Tully-Fisher relations of spiral and S0
galaxies}

\author{Michael J. Williams}{
  address={Sub-department of Astrophysics, Denys Wilkinson Building,
  Keble Road, Oxford OX1 3RH, UK}
  ,altaddress={ESO, Karl-Schwarzschild-Str. 2, D-85748 Garching bei
  M\"unchen, Germany} % additional visiting address
}

\author{Martin Bureau}{
  address={Sub-department of Astrophysics, Denys Wilkinson Building,
  Keble Road, Oxford OX1 3RH, UK}
}

\author{Michele Cappellari}{
  address={Sub-department of Astrophysics, Denys Wilkinson Building,
  Keble Road, Oxford OX1 3RH, UK}
}


\begin{abstract}
We construct mass models of 28 S0--Sb galaxies. The models have an
axisymmetric stellar component and a NFW dark halo and are constrained
by observed $K_{S}$-band photometry and stellar kinematics. The median
dark halo virial mass is $10^{12.8}\,M_\odot$, and the median dark/total
mass fraction is 20\% within a sphere of radius $r_{1/2}$, the intrinsic
half-light radius, and 50\% within $R_{25}$. We compare the Tully-Fisher
relations of the spirals and S0s in the sample and find that S0s are 0.5
mag fainter than spirals at $K_S$-band and 0.2\,dex less massive for a
given rotational velocity. We use this result to rule out scenarios in
which spirals are transformed into S0s by processes which truncate star
formation without affecting galaxy dynamics or structure, and raise the
possibility of a break in homology between spirals and S0s.
\end{abstract}

\maketitle

\section{Mass models}

In \cite{Williams:2009} we presented mass models for a sample of 28
edge-on early-type disk galaxies (S0--Sb). The models are composed of an
axisymmetric stellar component, based on observed $K_S$-band photometry
\cite{Bureau:2006} assuming a constant stellar mass-to-light ratio
$(M/L)_{K_S}$, and a spherical NFW dark halo \citep{Navarro:1997} of
mass $M_{\rm DM}$. The model parameters are constrained by solving the
Jeans equations assuming a constant anisotropy in the meridional plane
$\beta_z \equiv 1 - \sigma_z^2/\sigma_R^2$, which yields a prediction of the
second velocity moment, and comparing to observed stellar kinematics
\cite{Chung:2004}. These simple models are able to reproduce the wide
range of observed stellar kinematics, which extend to 2--3 effective
radii. 

The median $(M/L)_{K_S}$ for the sample is 1.09 (solar units) with an
rms scatter of 0.33. The median $M_{\rm DM}$ for the sample is
$10^{12.8}\,M_\odot$ with an rms scatter of 0.7\,dex. This is equivalent
to halo concentrations between 7 and 9. The mass models have a median
dark/total mass fraction of 20\% within a sphere of radius $r_{1/2}$,
the intrinsic half-light radius (approximately equal to 1.33\,$R_{\rm
e}$, where $R_{\rm e}$ is the projected half-light (or effective) radius
and 50\% within $R_{25}$. All but two models are consistent with being
maximal \cite{Sackett:1997}, although they were not constructed under
this assumption. Models without a dark halo are also able to reproduce
the observed kinematics satisfactorily in most cases, but the
improvements when halos are added are statistically significant.
Moreover, a preliminary comparison shows that the stellar mass-to-light
ratios of mass models without dark matter very significantly exceed the
predictions of stellar population models for plausible initial mass
functions, but this effect is significantly reduced for our preferred
mass models with dark haloes. We will use these results to measure the
normalization and universality of the IMF in more detail in the future.

\section{The Tully-Fisher relation}

\begin{figure}
\includegraphics{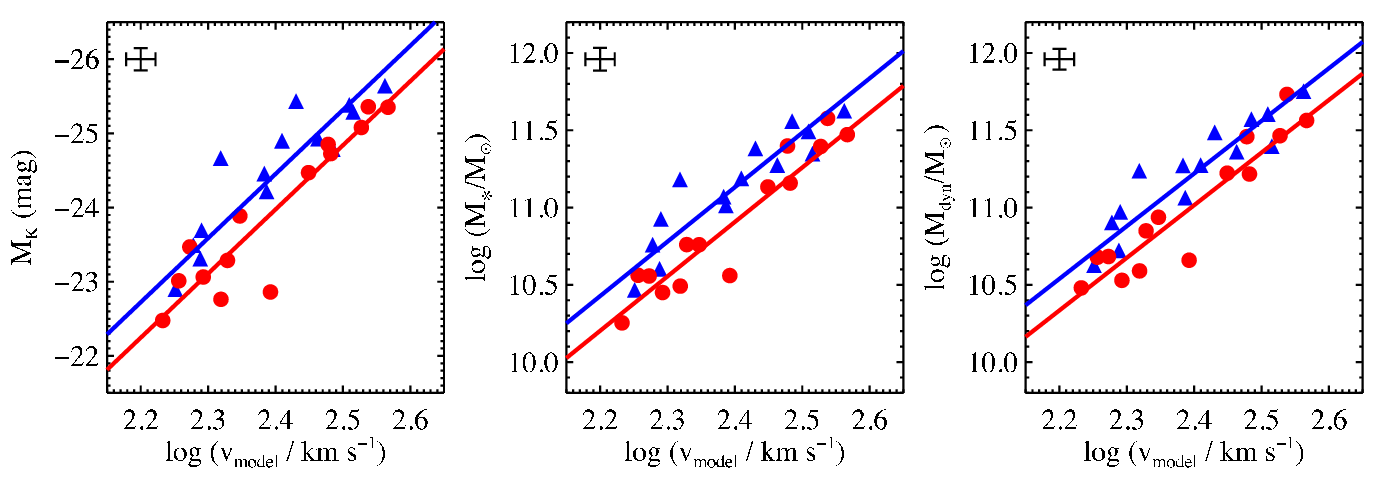}
\caption{Tully-Fisher relations of the sample of 28 disk galaxies, shown
as functions of $K_S$-band luminosity, stellar mass and dynamical mass.
Spirals are shown as blue triangles, S0s as red circles. Median error
bars are shown in the top left of each plot.}
\label{fig:tfr}
\end{figure}

Using the circular velocities of the models, we constructed Tully-Fisher
relations (TFRs) as functions of luminosity, stellar mass, and dynamical
mass for the S0s and spirals separately (Fig.~\ref{fig:tfr}). We find
that S0s are 0.5\,mag fainter than spirals at ${K_S}$-band for a given
rotational velocity. In stellar population synthesis models in which
star formation is truncated, this fading would take $\sim 1$\,Gyr, but
we know that the processes which form S0s began at earlier times
\cite{Dressler:1997,Fasano:2000}. We therefore rule out scenarios in
which spirals are transformed into S0s by an environmental or secular
process which simply truncates star formation, without affecting the
dynamics or structure of the galaxies. 

The offset of the S0 TFR could be explained by recent star formation in
S0s, but we find that the offset of the S0 TFR persists as a function of
both stellar and dynamical mass (S0s are 0.2\,dex less massive for a
given model circular velocity). The offset could therefore be explained
by a small (10--20\%) but systematic contraction or compactification of
spirals as they transform to S0s, an effect consistent with the
morphological dependence of the local size--luminosity relation
\cite{Courteau:2007}. This possibility is discussed in more detail in
our paper on the TFR \cite{Williams:2009a}.

\bibliographystyle{aipproc}   % if natbib is available
%\bibliography{/Users/mike/research/bibliography/bib}

\end{document}